# A self-consistent model for estimating the critical current of superconducting devices

V. Zermeño[1], F. Sirois[2], M. Takayasu[3], M. Vojenciak[4], A. Kario[1] and F. Grilli[1]

[1]Karlsruhe Institute of Technology, Karlsruhe, Germany

[2]Ecole Polytechnique Montréal, Montréal, Canada

[3]Massachusetts Institute of Technology, Cambridge, MA, USA

[4]Slovak Academy of Science, Bratislava, Slovakia

Email: victor.zermeno@kit.edu

## Abstract

Nowadays, there is growing interest in using superconducting wires or tapes for the design and manufacture of devices such as cables, coils, rotating machinery, transformers and fault current limiters among others. Their high current capacity has made them the candidates of choice for manufacturing compact and light cables and coils that can be used in the large scale power applications described above. However, the performance of these cables and coils is limited by their critical current, which is determined by several factors, including the conductor's material properties and the geometric layout of the device itself. In this work we present a self-consistent model for estimating the critical current of superconducting devices. This is of large importance when the operating conditions are such that the self-field produced by the current is comparable to the overall background field. The model is based on the asymptotic limit when time approaches infinity of Faraday's equation written in terms of the magnetic vector potential. It uses a continuous $E-J$ relationship and takes the angular dependence of the critical current density on the magnetic flux density into account. The proposed model is used to estimate the critical current of superconducting devices such as cables, coils, and coils made of transposed cables with very high accuracy. The high computing speed of this model makes it an ideal candidate for design optimization.

# Contents

# 1. Introduction

Nowadays, superconducting devices made of coated conductors such as cables and coils are being used for their high current capacity or their ability to create large magnetic fields within compact and light designs (see Figure 1). Estimating the critical current ($I_c$) of a superconducting device is of high importance for design and optimization. Although several manufacturers of superconducting wires and tapes provide general information for the $I_c$ characteristic of their conductors, in general this information lacks enough resolution regarding externally applied magnetic fields of small amplitude (comparable to the conductor's self-field) and their orientation. As pointed out in [1] the self-field due to transport current has a non-negligible contribution to the estimated $I_c$ characteristic when considering externally applied magnetic fields of small amplitudes. Furthermore, for applications such as cables and small coils, consideration of this self-field is necessary for accurate estimation of $I_c$ [2], [3]. For large devices operating in large fields, it is possible to obtain a good estimate of the critical current by simply calculating the magnetic field at a given current rate and by comparing it to the expected critical current of the tape at that field. This strategy, known as the load-line method, has been used for designing superconducting magnets and is well described in [4]. However, for smaller devices that operate in fields with amplitudes that are comparable to the self-field of the conductors, it is necessary to estimate the critical current of the device by considering the local self-field effects of the superconducting material. Recent improvements to the load-line method can be found in [5] and [6]. In these works, the critical currents of Roebel assembled coated conductors and transformers made with such cables are estimated. Although these approaches provided fairly precise models, being based on the load-line method, they still require an iterative and comparative process to estimate the critical current value and assume a uniform current distribution inside each conductor. A more precise, self-consistent, method needs to consider the critical current density dependence of the superconducting material on the local flux density. This dependence, denoted by $J_c(\boldsymbol{B})$, is usually obtained by measuring the $I_c$ of the tape at fields of different amplitudes and orientations. In general, this process involves solving an inverse problem to fit the parameters of the a predefined expression for $J_c(\boldsymbol{B})$. The reader is directed to [2], [7], [8] where this process is described in detail considering three different numerical approaches.

A previously published DC model [9] allowed estimating the critical current of an isolated single conductor. Other models [10], [11], [12] allowed already considering several conductors but they required an iterative estimation of the critical current or involved direct solution of the nonlinear $E - J$ relationship that is used to model the resistivity of the superconducting materials, hence requiring an involved numerical solution.

Estimation of the critical current of circular pancake coils has already been studied by means of the H-formulation of Maxwell's equations using an axisymmetric model [13]. The use of said strategy for critical current estimation requires solving a fully transient model starting from zero initial conditions. The transport current in the coils is then slowly ramped up in a staircase-like manner. Solution of this time dependent model results in a large computing time. More recently, an approach based on the critical state model was presented in [14]. Being history dependent, this model still requires performing computations starting from zero initial conditions and in general requires a much larger computing time than the DC models.

In the present work, we introduce a new DC model to directly estimate the critical current of superconducting devices such as cables, coils, and coils made of transposed cables with high accuracy. The model uses an auxiliary variable that is uniform in every domain (denoted by $P$) to account for the nonlinear $E-J$ relationship. This allows a very fast, robust and easy-to-implement method to estimate the critical current of devices composed of several conductors. Although most of the examples and validating experiments presented here consider the case of HTS coated conductors, the mathematical formulation is quite general and can easily be used to consider other types of superconducting wires without more effort than utilizing the appropriate parameters related to the materials used.

The mathematical description of the model is presented in section 2. In section 3, considerations regarding the determination of the $I_c$ of devices are discussed and several criteria matching different experimental setups are presented. Section 4 describes the extension of the model from section 2 to consider non-transposed conductors. Here, models for cables with ohmic termination resistances as well as for non-transposed twisted cables under an externally applied magnetic field are also presented. To prove the validity of the model, several comparisons against experimental results are carried out and presented in section 5. Section 6 presents further applications of the model to single tape coils, arrays of cables and to the case of a twisted cable under the influence of an externally applied magnetic field. A brief general discussion of the model, its validation and applications are given in section 7. The final section provides some concluding remarks.

## 2. Model Description

The model presented here derives from the asymptotic limit when $t \to \infty$ of the Maxwell-Faraday's equation $\nabla \times \mathbf{E} = -\partial \mathbf{B}/\partial t$. Expressing the magnetic flux density $\mathbf{B}$ in terms of the magnetic vector potential $\mathbf{A}$, as $\mathbf{B} = \nabla \times \mathbf{A}$, the electric field can be written as:

$$\mathbf{E} = -\frac{\partial \mathbf{A}}{\partial t} - \nabla V, \qquad (1)$$

where $V$ is the electric potential. When all excitations are DC sources, the limit $t \to \infty$ implies that $\partial \mathbf{A}/\partial t \to 0$, which physically means that the magnetic field inside the superconducting material has completely relaxed, and that $J$ has reached its steady state. This is obviously a simplified representation of the real physics occurring in a superconducting wire, in which the continuous DC voltage observed results from the presence of flux flow in flux percolation channels distributed along the length of the wire. That is, at the microscopic level, for a given

DC current near $I_c$, most sections of the wire exhibit a lossless critical state behavior, whereas a few sections are in the flux flow state, responsible for the DC voltage measured (see [15] for a detailed discussion about the inherent transport current physics near $I_c$). The 2-D model proposed in this paper must therefore be understood as one in which flux flow is averaged over the whole length of the wire. Even if this may appear to be a strong assumption, this model proved to behave much like macroscopic experimental results, as demonstrated later in this paper.

In the 2-D case considered here, equation (1) has only one scalar component, the one parallel to the direction of current flow. Therefore, we shall note as $E, J$ and $A$, the scalar components of $\mathbf{E}, \mathbf{J}$ and $\mathbf{A}$, respectively. With the DC condition $\partial \mathbf{A}/\partial t \to 0$, equation (1) can be rewritten simply as

$$\mathbf{E} = -\nabla V, \qquad (2)$$

where $V$ must be uniform over the cross-section of each conductor, otherwise the problem is no longer 2-D. Since the model considers infinitely long conductors, the voltage gradient $\nabla V$ is also uniform over the cross-section of each conductor. By extension, $E$ must also be uniform within each conductor, hence being equal to the voltage drop per unit length.

This work considers only materials for which a continuous injective function $f(\cdot)$ can be used to describe the relation between the electric field $E$ and the current density $J$, so that the $E - J$ relationship can be written as $E = f(J)$. The electric field at which the current density reaches its critical value $J_c(\mathbf{B})$ is denoted by the constant value $E_c$. A power law of index $n$ is commonly used to describe the $E - J$ relationship as follows [16]:

$$E = E_c \frac{J}{J_c(\mathbf{B})} \left|\frac{J}{J_c(\mathbf{B})}\right|^{n-1}. \qquad (3)$$

This relation can be inverted to become:

$$J = J_c(\mathbf{B}) \, P \qquad (4)$$

where

$$P = \frac{E}{E_c} \left|\frac{E}{E_c}\right|^{\frac{1}{n}-1}. \qquad (5)$$

Recalling that in the DC case, $E$ is constant over the cross-section of each conductor, $P$ is also constant. One can observe that when the current in a conductor reaches $I_c$, i.e. $J = J_c(\mathbf{B})$, then

$P = 1$ and $E = E_c$. The introduction of the variable $P$ allows avoiding the direct solution of the nonlinear $E - J$ relationship, hence providing a simpler problem that can be easily solved. Following an analogous principle, similar expressions to (4) can be obtained for other $E - J$ relationships.

For bundles of conductors carrying the same current, as it is the case for the different turns in coils or the strands in transposed cables, one constant $P_i \mid i \in \{1,2,\ldots,n_c\}$ for each of the $n_c$ conductors considered in the cross section is necessary. To impose a given current $I_a$ in the $i$-th conductor it suffices to choose $P_i$ so that:

$$I_a = \int_{\Omega_i} P_i J_c(\mathbf{B}) \, dxdy. \tag{6}$$

Here $\Omega_i$, denotes the domain corresponding to the $i$-th conductor. Being a constant, $P_i$ can be calculated as:

$$P_i = I_a / \int_{\Omega_i} J_c(\mathbf{B}) dxdy. \tag{7}$$

The voltage drop per unit length $E_i$ in the $i$-th conductor is obtained by substituting (4) in (3):

$$E_i = E_c P_i |P_i|^{n-1}. \tag{8}$$

To solve for the individual current densities and calculate the magnetic field, Ampere's law written in terms or the magnetic vector potential is used. Hence, with $\mathbf{B} = \mu \mathbf{H}$ one can write:

$$\nabla \times \frac{1}{\mu} \nabla \times \mathbf{A} = J. \tag{9}$$

Using (4), the current source term can be replaced yielding:

$$\nabla \times \frac{1}{\mu} \nabla \times \mathbf{A} = J_c(\mathbf{B}) \, P. \tag{10}$$

In the air/insulating regions, $P = 0$ while in the superconducting regions $P$ is given by the corresponding domain-wise uniform values for $P_i$ in (7). The problem now is reduced to solve equations (7) and (10) given appropriate boundary conditions related to each application of interest. The variable $P$ has the function of ensuring that each conductor carries the same net current $I_a$. Therefore, the cases of transposed cables, coils and coils made of transposed cables can be properly considered. One additional feature of the model presented here is that depending on the application under study, several criteria for $I_c$ can be considered. These criteria can be used to compare with experimental results where the voltage drop is typically measured across different locations of the device. In the following section, different $I_c$ criteria used for devices composed of several conductors are presented and discussed.

The model is easy to implement in both commercial and open source software packages for FEM modeling such as COMSOL [17] or FreeFem++ [18], [19]. For consistency, all the results presented in this work were calculated using COMSOL.

## 3. Selection of the $I_c$ criteria

In this section, we present different criteria to determine the critical current of superconducting devices. In general, such determination in a straight superconducting wire is quite straightforward: following the four point-measurement technique, a slowly ramping DC current is injected at the ends of the sample, and the voltage between two points on the sample between the current leads (or outside them) is read by means of voltage taps and recorded. When the voltage reaches a "critical" given value of the electric field $E_c$ (which in the case of HTS is usually set at 0.1 or 1 µV/cm), the corresponding current is called critical current. If one considers a slightly more complicated geometry, for example a pancake coil wound from a superconducting wire, things are less evident: usually the critical current of such coil is determined by measuring the coil's voltage end-to-end – one can call such critical current $I_c^{coil}$. It is however probable, especially if the coil is composed of many tightly packed turns, that, as a consequence of the generated self-field, some parts of the coil (for example the innermost turns) develop a critical electric field at currents lower than $I_c^{coil}$. This can be dangerous and lead to local over-critical situations, even at currents lower than $I_c^{coil}$. So, if one associates the critical current to the maximum current that a superconducting device can carry without risk of quench and/or damage, it is essential to understand how this limit is defined. This depends on the specific architecture of the device (e.g. cable made of transposed tapes, pancake coil, etc.). In the following, we address the definition of critical current for three cases of different complexity: transposed cables, single tape coils and coils wound from transposed cables.

### 3.1. Transposed cables

Consider the case of current transport in transposed cables such as Roebel type of cable [20], [21]. Since the cable is much longer than the transposition length of the strands, this transposition guarantees that the voltage along every strand is the same. However, some regions of the strands can locally experience larger electric fields than others. For instance, the region of the strands in the outermost part of the cable will experience an overall local magnetic field that is larger than those located in the center. This in turn will yield a local lower critical current density. In long cables, this effect is averaged as the transposition ensures that all strands experience the same electromagnetic conditions. Still, this averaging can mask potentially dangerous situations where the local electric field in certain regions could reach large values (for example due to non-uniform critical current in each strand or to local strand defects). If the voltage taps used for measurement are placed too far apart, this highly localized electric field can be overlooked in the measurement. In the 2D model described above, this leads to consider two different criteria for $I_c$ in a transposed cable composed of $k$ conductors:

- $I_c$ is the current at which the voltage drop per unit length $E_i$ has reached its critical value $E_c$ in *at least one conductor* i.e. $P_i = 1$. This can be obtained with the condition $\max_{i \in \{1,2...k\}} P_i = 1$. In what follows, we shall refer to this as the MAX criterion.
- $I_c$ is the current at which the *average* voltage drop per unit length $E_i$ has reached its critical value $E_c$. This can be obtained with the condition $\sum_{i=1}^{k} E_i/k = E_c$, or alternatively after use of (8), as $\sum_{i=1}^{k} P_i |P_i|^{n-1} = k$. In what follows, we shall refer to this as the AVG criterion.

### 3.2. Single tape coils

A situation similar to the one just described arises when considering coils wound using single tape. In self-field conditions, the innermost turn will typically be subjected to a higher electric field than the other turns and will reach its critical current sooner [14]. Measuring the voltage drop across the whole coil provides an averaged quantity as the voltage drop across individual turns is added, masking the higher local electric field in the innermost turn. Furthermore, consider the case of a uniform magnetic field being applied opposite to the central field of the coil. In this case, a large enough field might drive the outermost turn to reach its critical current faster than the innermost. In all cases, this leads to consider two different criteria for $I_c$ in a coil composed of $m$ turns with inner radius $r$ and turn spacing $s$:

- $I_c$ is the current at which the voltage drop per unit length $E_j$ has reached its critical value $E_c$ in *at least one turn* i.e. $P_j = 1$. This can be obtained with the condition $\max_{j \in \{1,2...m\}} P_j = 1$. By analogy with the transposed cable already considered, in what follows, we shall also refer to this as the MAX criterion.
- Considering that the length $l_j$ of the *j-th* turn of a coil is $l_j = 2\pi (r + s(j-1))$, $I_c$ is the current at which the sum of the voltage drop in each turn divided by the coil's length reaches the critical value $E_c$. With the total length of the simulated coil in the axisymmetric model given by $L = \sum_{j=1}^{m} l_j$, this condition can be stated as $\sum_{j=1}^{m} E_j l_j / L = E_c$, or alternatively after use of (8), as $\sum_{j=1}^{m} P_j |P_j|^{n-1} l_j = L$. In what follows, we shall refer to this as the SUM criterion.

### 3.3. Coils made of transposed cables

Finally, following from the previous discussion for transposed cables and coils, for a coil composed of $m$ turns made of a transposed cable with $k$ conductors, inner radius $r$ and turn spacing $s$, we can define the $I_c$ in three different ways:

- $I_c$ is the current at which the voltage drop per unit length $E_{i,j}$ has reached its critical value $E_c$ in *at least one conductor* of the cable used for winding the coil, i.e. $P_{i,j} = 1$. This can be obtained with the condition $\max_{i \in \{1,2...k\}, j \in \{1,2...m\}} P_{i,j} = 1$. By analogy with the previous considerations, in what follows, we shall also refer to this as the MAX criterion.

- Defining $\bar{E}_j = \sum_{i=1}^{k} E_{i,j}/k$ as the average voltage drop per unit length in the conductors of the $j$-th turn of a coil (alternatively, $\bar{E}_j = E_c/k \sum_{i=1}^{k} P_{i,j}|P_{i,j}|^{n-1}$), $I_c$ is the current at which $\bar{E}_j$ has reached its critical value $E_c$ for *at least one turn*. This can be obtained with the condition $\max_{j \in \{1,2...m\}} \bar{E}_j = E_c$, or alternatively after use of (8), $\max_{j \in \{1,2...m\}} \sum_{i=1}^{k} P_{i,j}|P_{i,j}|^{n-1} = k$. In what follows, we shall refer to this as the MAX-AVG criterion.

- Considering that the length $l_j$ of the $j$-th turn of a coil is $l_j = 2\pi (r + s(j-1))$, $I_c$ is the current at which the *sum of the averaged voltage drop* in each turn divided by the coil's length reaches the critical value $E_c$. If the total length of the coil is given by $L = \sum_{j=1}^{m} l_j$ and the average voltage drop per unit length in the conductors of the $j$-th turn of a coil is given by $\bar{E}_j = \sum_{i=1}^{k} E_{i,j}/k$, then this condition can be stated as $\sum_{j=1}^{m} \bar{E}_j l_j = E_c L$, or alternatively after use of (8), as $\sum_{j=1}^{m} \sum_{i=1}^{k} P_{i,j}|P_{i,j}|^{n-1} l_j = L\,k$. In what follows, we shall refer to this as the SUM-AVG criterion.

## 4. Model extension to non-transposed cable conductors

After some small modifications, the model presented in the previous sections can also be used for analysis of non-transposed cable conductors. In this section, three interesting applications of the present model to these conductors are treated and tools are presented to estimate their critical currents.

### 4.1. Non-transposed straight cables

The critical current of non-transposed straight cables can be estimated by considering that the net current repartition among the strands is not necessarily uniform as in the case of transposed cables. Furthermore, since the strands are connected at both ends, current sharing among all strands through the current leads is possible. In general, cable strands that are closer to the cable's axis will experience a lower magnetic field than the ones in the outer layers. Therefore, the strands in the external layers will have a lower critical current value than their internal counterparts. As a consequence, in the case of non-transposed cables, the critical current of the cable is obtained when the current in each individual strand reaches its corresponding critical value. Considering the mathematical formulation presented in the model description of section 2, this condition can be implemented by considering $E_i = E_c$ and consequently $P_i = 1$ for $i \in \{1,2,\dots,n_c\}$ in each of the $n_c$ strands of the cable and $P = 0$ in the air/insulation domains in equation (10). It is important to mention that such simplification of the model presented here converges to the previously reported approach described in [2], [10] where all conductors are assumed to be in the critical state and therefore, not necessarily carrying the same current each.

## 4.2. Cables with ohmic termination resistances

Besides the self-field effect and the variation on the individual $I_c$ of each conductor, ohmic termination resistances in superconducting cables are largely responsible for the uneven current distribution in the cable strands. These problems have been already addressed in [22] for non-transposed cables such as the Twisted Stacked-Tape Cable (TSTC) (see Figure 1b). There, three different models are proposed and validated via comparison with experimental data. The models presented offer a palette of options to study the current sharing among the strands. However, in said work, the most accurate models presented – accounting for both the termination resistance and the critical current density dependence on the magnetic field – require the use of a transient approach that involves a long computation time. In this section we present a new method to account for the ohmic termination resistances in superconducting cables while considering the critical current density dependence on the magnetic field in a simpler and faster model than those presented in [22].

The voltage drop per unit length $E_i$ in the i-th conductor will now have an additional contribution $\bar{E}_i^{tr}$ given solely by the termination resistances. This contribution can be easily considered by modifying (8) so that $E_i$ can be expressed as:

$$E_i = E_c P_i |P_i|^{n-1} + \bar{E}_i^{tr} \tag{11}$$

where:

$$\bar{E}_i^{tr} = \bar{R}_i I_i. \tag{12}$$

Here, $\bar{R}_i$ is the scaled resistance (in $\Omega/m$) of the termination at its ends. The scaling in the values for $\bar{R}_i$ allows considering a "per unit of length" model. The current in the i-th superconducting tape denoted by $I_i$ can be calculated as:

$$I_i = P_i \int_{\Omega_i} J_c(\boldsymbol{B}) \, dxdy. \tag{13}$$

Substituting (12) and (13) into (11) yields:

$$E_i = E_c P_i |P_i|^{n-1} + \bar{R}_i P_i \int_{\Omega_i} J_c(\boldsymbol{B})\, dxdy. \tag{14}$$

Since all the strands – together with their corresponding termination resistances – are connected in parallel, the voltage drop per unit length across them is the same. Therefore, we can write $n_c - 1$ equations of the form:

$$E_c P_i |P_i|^{n-1} + \bar{R}_i P_i \int_{\Omega_i} J_c(\boldsymbol{B}) dxdy = E_c P_{i+1}|P_{i+1}|^{n-1} + \bar{R}_{i+1} P_{i+1} \int_{\Omega_{i+1}} J_c(\boldsymbol{B}) dxdy, \tag{15}$$

$$i \in \{1,2,\ldots,n_c - 1\}.$$

An additional equation is obtained by considering the total current in the cable $I_a$ as an input parameter. For a cable with $n_c$ strands this relation is given by:

$$I_a = \sum_{i=1}^{n_c} P_i \int_{\Omega_i} J_c(\boldsymbol{B})\, dxdy. \tag{16}$$

In the air/insulating regions, $P = 0$ while in the superconducting regions $P$ is given by the corresponding domain-wise uniform values for $P_i$ obtained from the solution of the nonlinear system given by equations (15) and (16). The problem now is reduced to solve equations (10), (15) and (16) given appropriate boundary conditions related to each application of interest. The critical current value of the cable can be estimated with both the MAX and the AVG criteria as discussed in Section 3.1. It is then required to fulfill the condition $max_{i \in \{1,2\ldots n_c\}} P_i = 1$ for the MAX criterion and $\sum_{i=1}^{n_c} P_i |P_i|^{n-1} = n_c$ for the AVG criterion.

### 4.3. Twisted cables under an externally applied magnetic field

Estimation of the critical current of twisted cables such as the TSTC in self-field conditions can be realized in a similar fashion to the non-twisted case described in 4.1. The helical symmetry of these cables allows estimating their critical current capacity by analyzing a single cross section plane. However, the application of an externally applied field breaks the helical symmetry and a cross section plane analysis does not provide the necessary conditions to estimate their critical current. In general, a 3D analysis should be able to accurately estimate the critical current in such cables. However, 3D simulations are in general computationally expensive and also require longer implementation times [23]. An alternative is to consider several cross sectional planes located in specific sections of the cable. In what follows, these cross sectional planes will be referred to as slices. In the case of cables assembled using stack-like arrays of tapes, such as the TSTC, two slices can be considered concerning the cases where the externally applied is either parallel or perpendicular to the tapes in the cable. More slices considering intermediate cases can also be included to increase accuracy.

Accounting for inter-strand current flow requires a detailed knowledge of the contact resistance along the cable length that is in general difficult to measure or estimate. Furthermore, such case leads to a truly 3D problem that require taking into account the current that flows across strands of the cable. Such case will be dealt with in future analysis and in the present work focus will be made in cables designs and conditions in which no inter-strand current flow is expected (electrically insulated superconducting strands).

In what follows, a cable with a layout similar to the TSTC will be studied [24]. For simplicity and without loss of generality, we study here the case where two slices are considered. These slices will be denoted by the computational domains $\Omega_\alpha$ and $\Omega_\beta$ (see Figure 11). Both cross sections are purposely chosen so that the tapes are parallel to the external field in one domain while they are perpendicular in the other. The governing equation is given by (10) as described in section 2. In the air/insulating regions, $P = 0$ while in the superconducting regions $P$ is given by the corresponding domain-wise uniform values for $P$ in the $2\,n_c$ superconducting sub-domains considered (one for each conductor in each slice). In such a cable design and regardless of the external field orientation or magnitude, a strand must carry the same net current along the whole length of the cable. Therefore, when modeling by means of two or more 2D slices it is necessary to constrain the current in a given strand so that it is the same in every slice model. This can be achieved using the proposed modeling strategy by constraining the current in a given tape to have the same amplitude in each computational domain. For such purpose, the following set of $n_c$ equations suffices:

$$P_{\alpha,i} \int_{\Omega_{\alpha,i}} J_c(\mathbf{B})\, dxdy = P_{\beta,i} \int_{\Omega_{\beta,i}} J_c(\mathbf{B})\, dxdy \ \forall\ i \in \{1,2,\ldots,n_c\}. \tag{17}$$

The critical current condition can be modeled by observing that the current in each tape will be limited by at least one of the slices. This means that once a tape has reached its critical current value in a slice, the current in that tape will be limited in all other slices. Again, both the MAX and the AVG criteria as discussed in Section 3.1 can be used for estimating the critical current. It is then required to fulfill the condition $\max(P_{\alpha,i}, P_{\beta,i}) = 1\ \forall\ i \in \{1,2,\ldots,n_c\}$ for the MAX criterion and $P_{\alpha,i} + P_{\beta,i} = 2\ \forall\ i \in \{1,2,\ldots,n_c\}$ for the AVG criterion.

The problem now is reduced to solve equations (10) and (17), given appropriate boundary conditions and critical current criteria related to each application of interest. Finally, the critical current in the cable $I_c$ can be calculated from a single slice as:

$$I_c = \sum_{i=1}^{n_c} P_{\alpha,i} \int_{\Omega_{\alpha,i}} J_c(\mathbf{B})\, dxdy. \tag{18}$$

## 5. Model Validation

As a way to validate the proposed model, comparison with experimental data was carried out by considering a 10-strand Roebel cable, five pancake coils wound using that same Roebel cable and a 4-tape TSTC in which the effect of termination resistance was considered. In all cases the experiments were performed in liquid nitrogen at 77 K. The Roebel cable was made by punching 5.5 mm wide strands out of a 12 mm wide tape and the TSTC was assembled by stacking 4 mm wide tapes. Both coated conductor tapes were manufactured by Superpower, Inc. [25].

As the normal metal layers in the CC architecture (in Superpower tape: Cu, Ag and Hastelloy) have an electrical resistivity that is several orders of magnitude higher than that of the superconducting material, only the superconducting layers were taken into account for simulation. In all cases, the dependence of the critical current density $J_c$ with respect to the magnetic flux density $\boldsymbol{B}$ was considered. In practice, this required solving the related inverse problem of finding a $J_c(\boldsymbol{B})$ such that experimental data for $I_c(\boldsymbol{B})$ could be numerically reproduced. For the Roebel cable, said relation was the one reported in [2]. For the TSTC, the expression for $J_c(\boldsymbol{B})$ reported in [22] was used. Brief descriptions of each cable are provided in the following subsections. However, the reader is directed to both [2] and [22] for further information regarding specific details of each cable.

### 5.1. Roebel cables

As a first test case, the problem of estimating the critical current of a Roebel cable is considered. The Roebel cable is composed of 10 strands punched from 12 mm wide coated conductor tape. The strands have 5.5 mm wide straight sections and a transposition length of 125.8 mm. Experimentally, the $I_c$ was determined by measuring the voltage drop along the length of a 4.5 m long cable. Since the transposition length of the cable is much shorter than its length, this corresponds to the AVG criterion described above as all strands experience the same electromagnetic conditions. One voltage tap pair per strand was placed in the cable, yielding uniform results [26]. Using a criterion of 1µV/cm it was found experimentally that the cable had a $I_c$ of 1002 A. For the AVG criterion our model yielded a value of 1038 A. The error with respect to experimental data is small (below 4%). This close estimate is remarkable considering that the model considers only the geometrical properties of the cable and the $J_c(\boldsymbol{B})$ dependence which was obtained for a single 12 mm tape [2],[26]. The numerical simulation using the MAX criterion yields a $I_c$ of 1005 A. This implies that for the transport current of 1038 A, at least one strand in the cable cross-section has already passed its local critical current. However, this event has effectively been averaged due to the transposition of the strands. Although in this case both the MAX and the AVG criteria provided similar estimates, for other configurations they could provide much larger differences. In general the

MAX criterion provides more conservative estimates than the AVG. It is important to consider this issue since large local overcritical currents might compromise the safe operation of the device.

## 5.2. Coils made of Roebel cables

As a second test case, coils made of Roebel cables were considered. The coils were wound using the Roebel cable described in the previous section. Five different coil were assembled from the same Roebel cable as reported in [3], [26]. These corresponded to coils with different inter-turn gaps. The gaps considered were 0.1, 1.4, 4, 10 and 20 mm. Figure 7 shows the critical current of the coils measured experimentally and the corresponding estimates given by the three criteria discussed in section 3.3. The critical current of the coils was measured experimentally by placing voltage taps along the total length of the superconducting coil and considering a critical electric field $E_c$ criterion of 1μV/cm. This voltage taps arrangement is consistent with the SUM-AVG criterion described in section 3. For the whole range of inter-turn separation gaps considered, the model predicts critical currents that are within 4% to 6% from the experimentally measured values. The critical current of the coils shows a monotonic increase with the size inter-turn separation gap. This is a consequence of the critical current dependence upon the local magnetic flux density. For coils with smaller distance between windings, the magnetic field produced by the coil is larger (see Figure 6) and therefore the $I_c$ of the coil will be lower than for coils with larger spacing. Estimates provided by the other $I_c$ criteria are also given for reference. In particular, it is notable that the critical current estimated by the – most conservative — MAX criterion is only 7% or 8% lower than the SUM-AVG criterion estimate. It is important to understand how the *local* critical current values of the cable at every turn relate to the SUM-AVG criterion, which is consistent with experimental measurements. For instance, the turns in the central region of the coil are exposed to a significantly larger magnetic field than the rest (see Figure 6).Therefore, these turns will have a lower local critical current value. Since the SUM-AVG criterion averages the voltage drop contribution of all the strands in all the turns, it is possible for a particular turn to reach an overcritical current without this being detected by the experimental setup. For this purpose, we calculate the normalized current per turn $\hat{I}_\tau = I_c/I_{c,\tau}$. Here $I_c$ is the critical current of the coil, estimated with the SUM-AVG criterion (see Figure 7) and $I_{c,\tau}$ is the critical current of the turn $\tau$. The results are shown in Figure 8 for the five Roebel coils studied. One can note that all coils present local overcritical currents only in their first 2 turns. The following turns present a decrease in $\hat{I}_\tau$. For coils with smaller inter-turn separation gaps, $\hat{I}_\tau$ increases again towards the last turns of the coil. A second local maximum is reached in the last turn of the coil. Just like in [26], a magnetic field dependent critical current density $J_c(\boldsymbol{B})$ was used to model the self-field effects. This effect can be observed by comparing Figure 6a and Figure 8. For the coil with 4 mm spacing, the minimum value for $\hat{I}_\tau$ is observed in the 7th turn (see Figure 8). This turn is indicated with a white arrow in Figure 6a. It is easy to see that the local minimum in $\widehat{I_\tau}$ corresponds to a turn in a region of low local magnetic field, hence evidently displaying the effect of the self-field.

### 5.3. Non transposed cables with ohmic termination resistances

As a final validation example, and to show the flexibility of the model described, the case of non-transposed cables with ohmic termination resistances is also considered. As mentioned in section 4.2, this modeling problem has already been addressed in [22] by means of other modeling tools that either disregard the dependence of the critical current density on the magnetic flux density or that require solving transient problems, which in general require longer computational times than stationary models. In this section we apply the model described in section 4.2 to the experimental test case presented in [24] and analyzed in [22]. Although said experiment is described in detail in the referred publications, for the sake of completeness, a brief description in given below.

The experimental test case analyzed here considers a 60 cm long straight stacked-tape cable composed of four stacked superconducting tapes, each tape being 4 mm wide. A separation of 100 μm between the superconducting layers is assumed to account for the non-superconducting materials in the tapes. A schematic of said cable is presented in Figure 9. The cable has resistive terminations at both ends. A slow current ramp (duration of 120 s) was applied to the cable until a net transport current of 275 A was reached. For reference, numbers 1 through 4 are used to denote the tapes in the cable (see Figure 9). As described in [22], values for the resistance of the terminations were obtained from experimental results. In the same manner, to account for the dependence of the critical current density on the local magnetic field, the same $J_c(\boldsymbol{B})$ relation reported in [22] was used here.

Figure 10 compares the simulations with experimental data. Here, the current in each tape of the cable is plotted as a function of the total current. It is important to note here that the model provides a good qualitative and quantitative representation of the experimental data over a large range of total current values.

The estimated critical current values given by the MAX and AVG criteria are respectively 225 A and 247.5 A. Although in [22] the critical current of the cable was not directly calculated, the critical current of each tape while *in the cable* configuration was estimated from experimental measurements. These values were respectively 65.3, 65.3, 64.4 and 55.4 A for tapes 1, 2, 3 and 4 by using a 1 μV/cm criterion. In the same way, the critical current values estimated using the model presented in this work are respectively 65.7, 64.5, 63.8, 56.3 A (an average error of 1.1% with respect to experimental data). These last current values, along with their corresponding cable current, are marked with large dots as a reference in Figure 10. It is easy to see that tape

1 reaches its critical current value at a lower cable current than the other tapes, therefore providing the critical current value related to the MAX criteria.

Since the critical current of each tape is not reached at the same time, once the current in a tape approaches its critical value, any additional current applied to the cable will mostly flow in the tapes that have not yet reached their corresponding critical current value. This means that even when most tapes have already reached their respective critical values, the additional current supplied to the cable will have little influence on the individual transport current of these tapes. This can be observed in Figure 10 in the curves of tapes 1, 3 and 4: very little changes in their transport currents are observed before their individual critical current values are reached. On the other hand, when the current in the last tape (#2) approaches its critical value (corresponding to a 251.4 A cable current), the transport current in all tapes increases again to accommodate any additional current supplied to the cable. It is important to note that the AVG criterion relates to the average voltage drop per unit length. Although this criterion cannot be directly inferred from Figure 10, such critical current value should be lower than the one at which all the tapes have reached (or passed) their individual critical current value (251.4 A). This is in good agreement with our AVG criterion estimate of 247.5 A.

## 6. Further Applications

As a way to show the power of the model and its capability to perform more involved studies, several applications of interest are discussed below. Again, just like in the previous section, in all cases the device is assumed to be in a liquid nitrogen bath at 77 K.

### 6.1. Inductive and anti-inductive stacks of transposed cables

As a complementary study – and to show the power of the method here presented – the cases of both inductive and anti-inductive infinite stacks of Roebel cables are considered as shown in Figure 2. Studying such arrays is of importance when considering Roebel cables to wind-up racetrack coils for motors or generators, or as part of fault current limiters. A parametric sweep spanning several values for the separation gap between the cables in an infinite array was carried and the critical current in each configuration was estimated using both MAX and AVG criteria. The results are presented in Figure 3. Estimates based on the MAX and AVG criteria are shown in red (dashed line) and blue (solid line) respectively. For reference, horizontal lines corresponding to the previously estimated values of $I_c$ for an isolated Roebel cable predicted by each criterion are plotted. Overall, both criteria yielded similar qualitative estimates for $I_c$ for the different separation distances. In general, the AVG criterion presented higher estimates than those of the MAX criterion. However, for small separations in inductive stacks, both MAX and AVG criteria converged to the same critical current of 599 A. For an inductive configuration with no separation, the distance between the strands of one cable and the following are the same as the distance between the strands within the same cable. In a way, by reducing the separation among the cables in the inductive stack the system looks more like two infinite stacks of equally spaced strands. In that case, all strands will experience the same electromagnetic properties and behave in the same way. Therefore both the MAX and AVG criteria are equivalent in such conditions. Finally, it is interesting to note that, as the separation between the turns increases, both the inductive and anti-inductive configurations converge to the previously estimated values of $I_c$ for an isolated Roebel cable predicted by each criterion. This is easily understood since for large separations the cables will experience electromagnetic conditions similar to the ones of isolated cables regardless of whether the array is inductive or anti-inductive. The numerical model proved to be very fast as this parametric sweep was carried out in less than 5 minutes using a laptop computer.

## 6.2. Coils wound using single tapes under applied magnetic field

The modeling technique presented in this work can also be used to estimate the critical current of coils in several non-trivial conditions. For instance, the conductors in a coil to which an external magnetic field is applied will have a different critical current than those in self-field conditions. In self-field conditions, the innermost turn typically becomes critical before the others. On the other hand if the externally applied field is in the direction opposite to the dipole's moment, a large enough field can drive the outer turn to reach its critical current faster than the innermost.

To analyze such configuration, a circular pancake coil composed of 50 turns made of 12 mm wide CC tape and a separation gap of 0.4 mm between turns is considered. Besides the case of transport current alone, the case of transport under applied magnetic field (both parallel and antiparallel to the dipole moment) is also studied. The same material properties used for analyzing the Roebel cable and coils in sections 5.1 and 5.2 were used here. Figure 4 shows the magnetic flux density in the considered coil for several values of interest for the externally applied magnetic field. In this section, positive and negative signs are used to indicate that the external field is respectively parallel or antiparallel to the coils' own magnetic moment. The limiting turns of the coil are pointed by the white arrows in Figure 4. Negatively-oriented fields of magnitude larger than 65 mT (considering the sign convention from above) cause the outermost turn to limit the overall current in the device. Alternatively, the innermost turn reaches its critical current value for other applied fields. It is interesting to note that for the particular case of an applied magnetic field of -65 mT, both the innermost and outermost turns reach their critical current value at the same time.

The $I_c$ values of the coil corresponding to different applied fields are shown in Figure 5. Both MAX and SUM criteria show that the maximum critical current happens for an applied field of -65 mT. However, at this field, the MAX criterion estimates an $I_c$ of 138 A, while the SUM criterion yields 145 A. Also, in Figure 5 (second vertical axis), the contribution from the coil to the total magnetic field is shown. This was obtained by evaluating the magnetic field at the center of the coil and subtracting the externally applied field from it. Being proportional to the current supplied, the magnetic field produced by the coil also presents a maximum for an applied field of -65 mT. The latter consideration is important in magnet design, as it allows considering not just the critical current of a coil while under externally applied field, but also the actual contribution it makes to the overall field.

Following a similar procedure, larger designs such as solenoid magnets or stacks of pancake coils can be simulated. This strategy allows analyzing complex magnet systems in a self-consistent manner, hence providing a valuable tool for design and optimization.

### 6.3. TSTC conductor under externally applied field.

Following the method described in section 4.3 (see also Figure 11), the case of a 40-tape TSTC under externally applied uniform magnetic field was considered. Such cable design and electromagnetic conditions are of interest in superconducting magnet design. For simplicity and to focus on the effect of the externally applied field, in this example, the termination resistance distribution is assumed to be uniform. Figure 12 shows the critical current of the cable under several externally applied uniform magnetic field values as estimated by the AVG and MAX criteria. One can note that for low values of the externally applied field, both criteria provide similar estimates. The reason for this is that externally applied fields of small amplitudes (when compared to the cable self-field) have little influence on the electromagnetic response of a cable. Therefore, for such low fields, the particular orientation of the magnetic field with respect to the tapes in the cable has little influence on its critical current. Hence, the $P$ values associated to a given tape are similar in all the slice models and the AVG criterion is similar to the MAX criterion. As the externally applied field increases, the $P$ values associated to a given tape are no longer similar in every slice model. Hence both MAX and AVG criteria provide different estimates. This change in behavior is clearly observed in Figure 12 at fields above 100 mT. Just like before, the MAX criterion provides more conservative estimates for the critical current. For example, for a field of 100 mT, the MAX and the AVG criteria estimate a critical current of 1459 A and 1564 A respectively.

## 7. Discussion

The estimated critical current values given by the model presented are in remarkably good agreement with experimental results in a variety of setups, including cables and coils. Small discrepancies can easily be understood considering several factors including: longitudinal inhomogeneity in the properties of the superconducting tapes, damage to the conductor during device fabrication (cabling process, tape punching, soldering, coil winding) or during repeated experimental testing (manipulation or thermal cycling) among others.

The model presented in this work allows obtaining an accurate estimate for the critical current of superconducting devices both in self-field and under an externally applied field. In systems that experience large background fields, rough estimates of their critical current can be obtained by use of the load-line method. This is a fair assumption in large devices or devices under a large background field where the field within a given conductor can be considered to be uniform. However, in the case of small devices, such as cables or coils, especially in self-field conditions, a fair estimate of the critical current cannot be achieved with such simple approximations, as the field within a given tape is likely to be far from uniform.

Although several criteria to estimate the critical current of devices were considered, the MAX criterion provides a more conservative estimate for the critical current than the others. Since this criterion considers the region of the device that reaches its critical value first, it can be used to identify "current bottlenecks". This information can be used in the design of the device to prevent inadvertently reaching local critical current values during normal operating conditions as these could damage the device. In a similar fashion identification of these "current bottleneck" regions allows evaluating the best locations for placing sensors of quench protection systems. For example, in the most obvious case of coils in self-field conditions, it would be advisable to measure the voltage drop across the innermost turn as compared to measure the voltage drop across the whole coil. For other devices than such a simple coil, regions of the current-bottleneck are not obviously identified as they might depend on their operating conditions. In such cases an analysis using the MAX criterion can provide a good estimate of the most vulnerable regions.

## 8. Conclusion

In this work, a model to calculate the critical current in superconducting power devices considering a self-consistent approach has been presented. The model follows from the asymptotic limit when time approaches infinity of Faraday's equation written in terms of the magnetic vector potential. The electromagnetic properties of the superconducting materials are considered by a continuous $E - J$ relationship and a critical current density that depends on the magnetic flux density. The model's calculations were compared against experimental results for a variety of setups including cables and coils. In all the cases considered, very good agreement with experiments was found. Several additional applications of the model were also presented to show its strength, flexibility, high speed and capability to perform involved studies. A brief discussion is given on how this model can be used to identify the current limiting regions due to the architecture of the superconducting devices. Such information is important on the design of built-in diagnostics and voltage tap positioning to identify and quickly react to large voltage surges appearing due to self-field effects that could damage the device. As a final note, it is worth mentioning that the model presented here is quite general and can be used to study any superconducting wire by simply choosing appropriate material parameters.

## 9. Acknowledgements

This work was partly supported by the Helmholtz Association (Young Investigator Group Grant VH-NG-617).

**Figures**

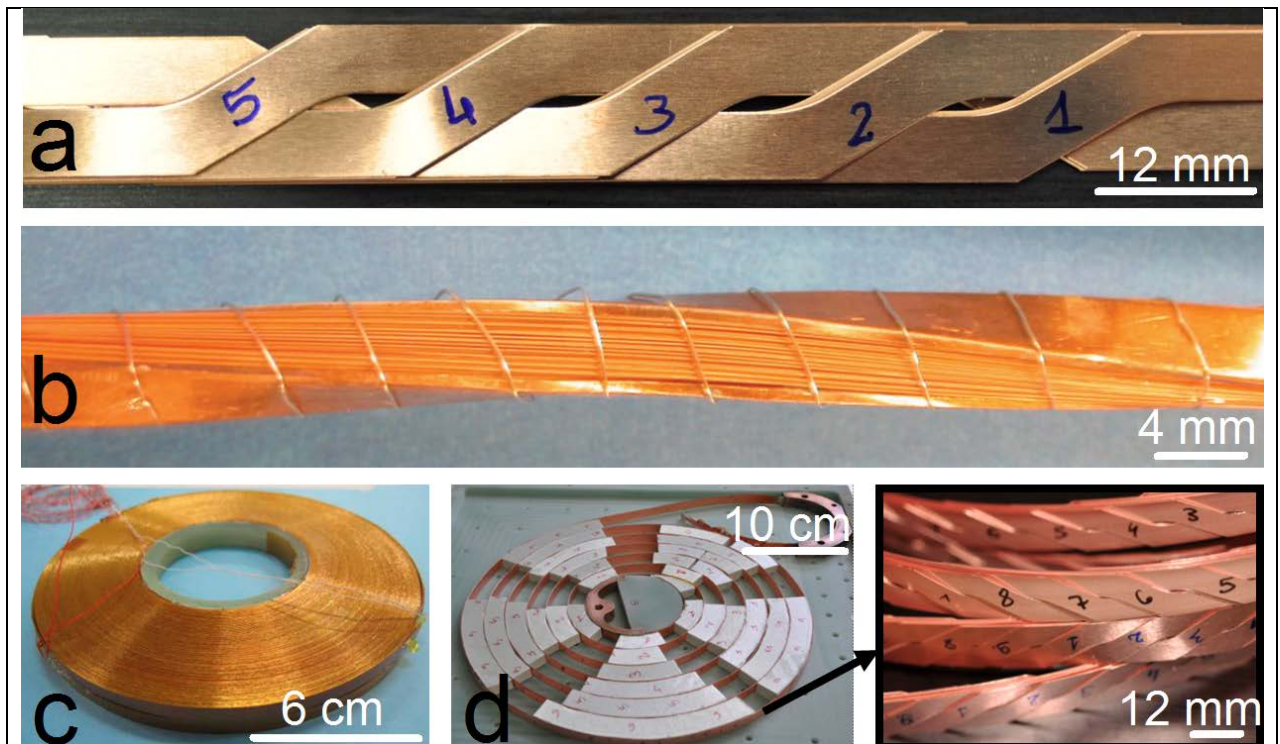

Figure 1. Cables and coils considered in this work. Roebel cable assembled with CC tapes (a). TSTC conductor assembled with CC tapes (b) [27]. Circular "pancake" coil assembled with a single CC tape (c) [28]. Circular "pancake" coil assembled a Roebel cable (d).

%% This line is just a spacer %%

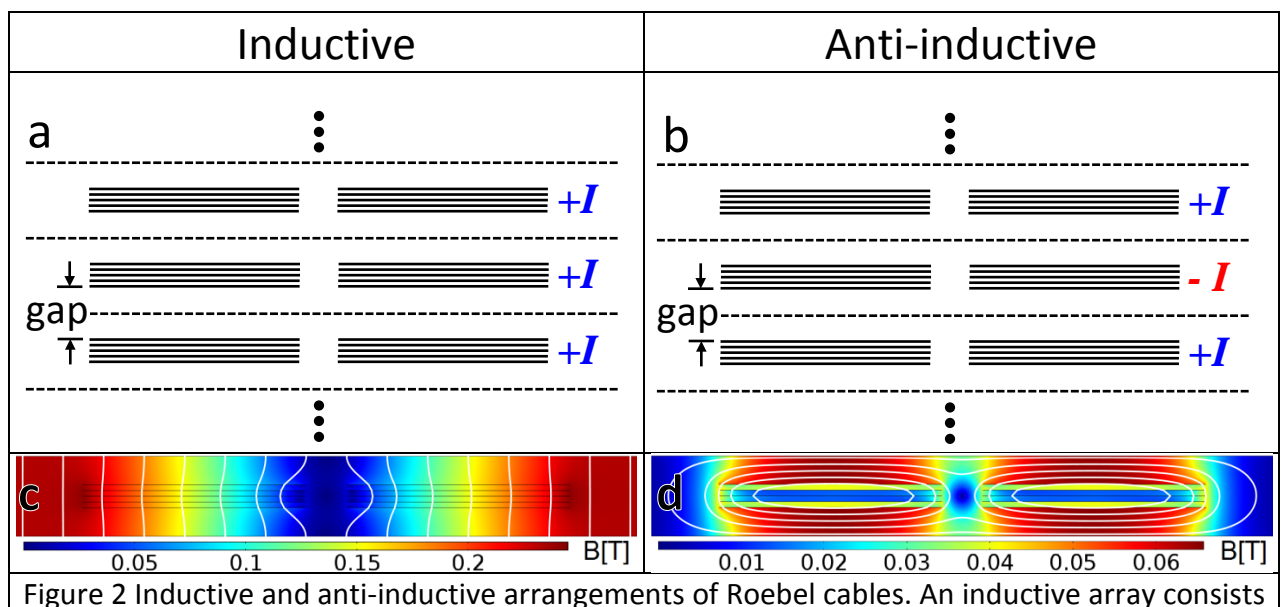

Figure 2 Inductive and anti-inductive arrangements of Roebel cables. An inductive array consists

of cables transporting a current of the same amplitude in the same direction (a). An anti-inductive array consists of cables transporting a current of the same amplitude but opposite directions (b). The amplitude of the magnetic flux density in the cross section of a single Roebel cable is also shown in (c) for the inductive and in (d) for the anti-inductive configurations.

%% This line is just a spacer %%

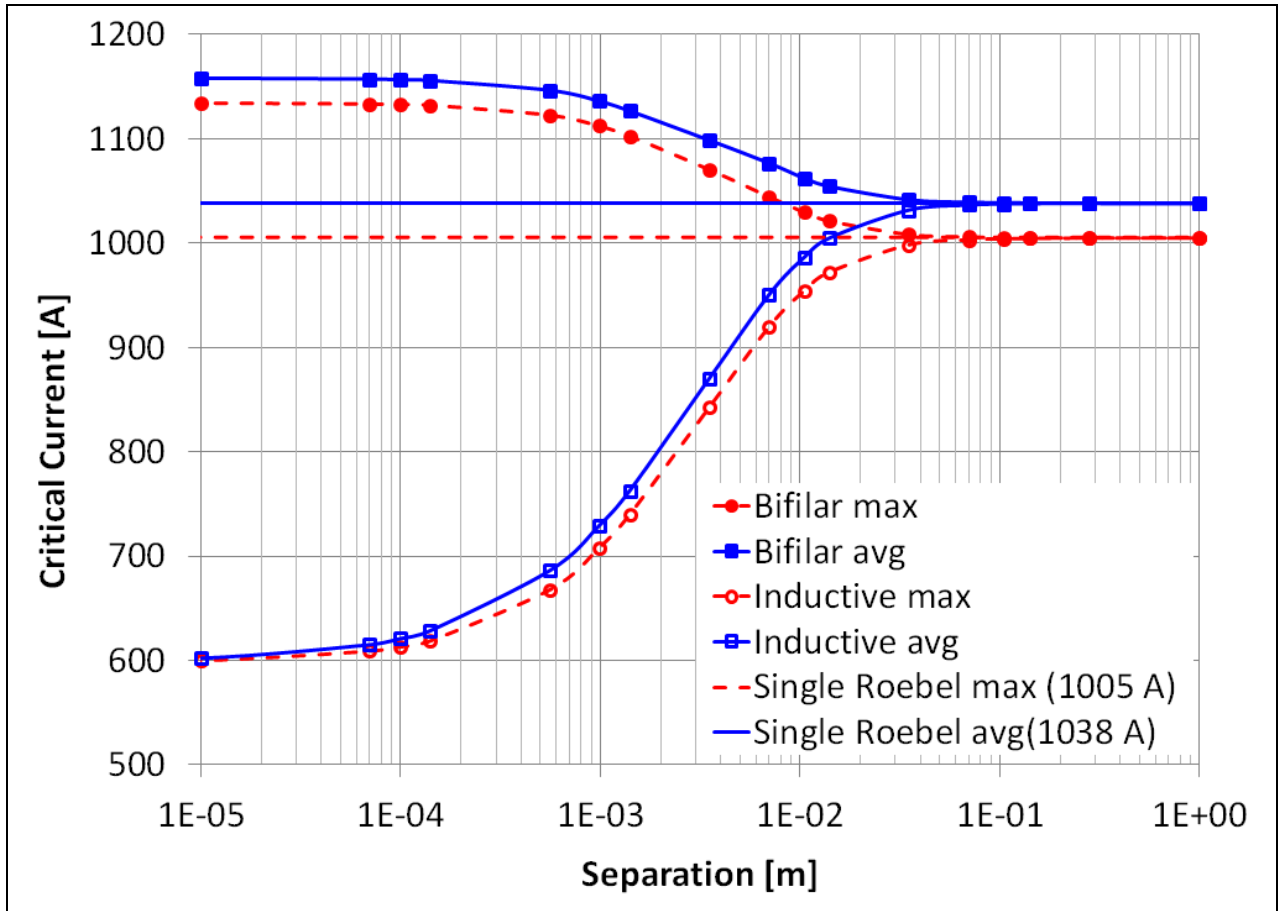

Figure 3 Critical current of inductive and anti-inductive arrays of Roebel cables with different gap separation as estimated by both the MAX and AVG criteria. For reference, critical current estimates for an isolated Roebel cable are also provided (horizontal lines).

%% This line is just a spacer %%

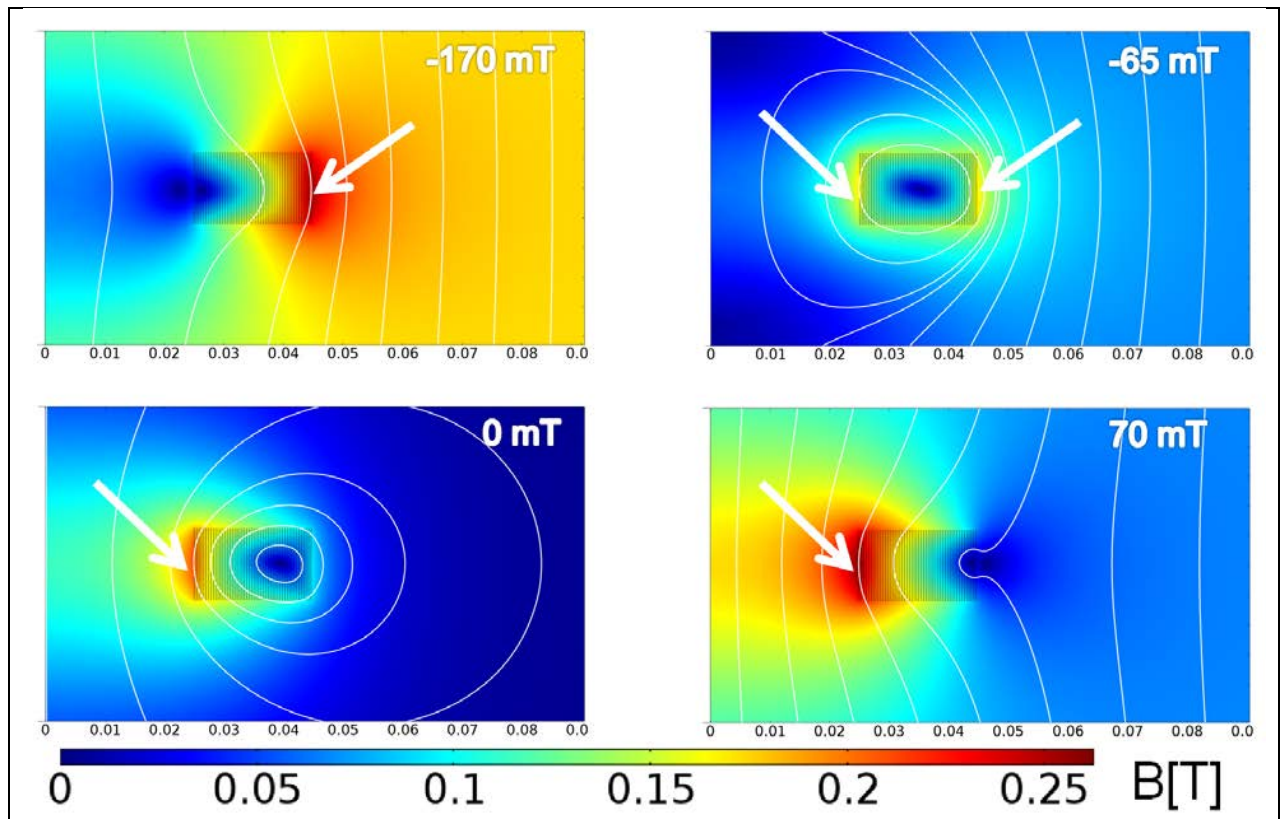

Figure 4 Magnetic flux density in pancake coils under externally applied field. The labels indicate the externally applied magnetic field. Positive and negative values denote external fields that are respectively parallel or antiparallel to the coils' magnetic moment. White arrows point to the current limiting turn in the coil i.e. the turn that reaches its local critical value first.

%% This line is just a spacer %%

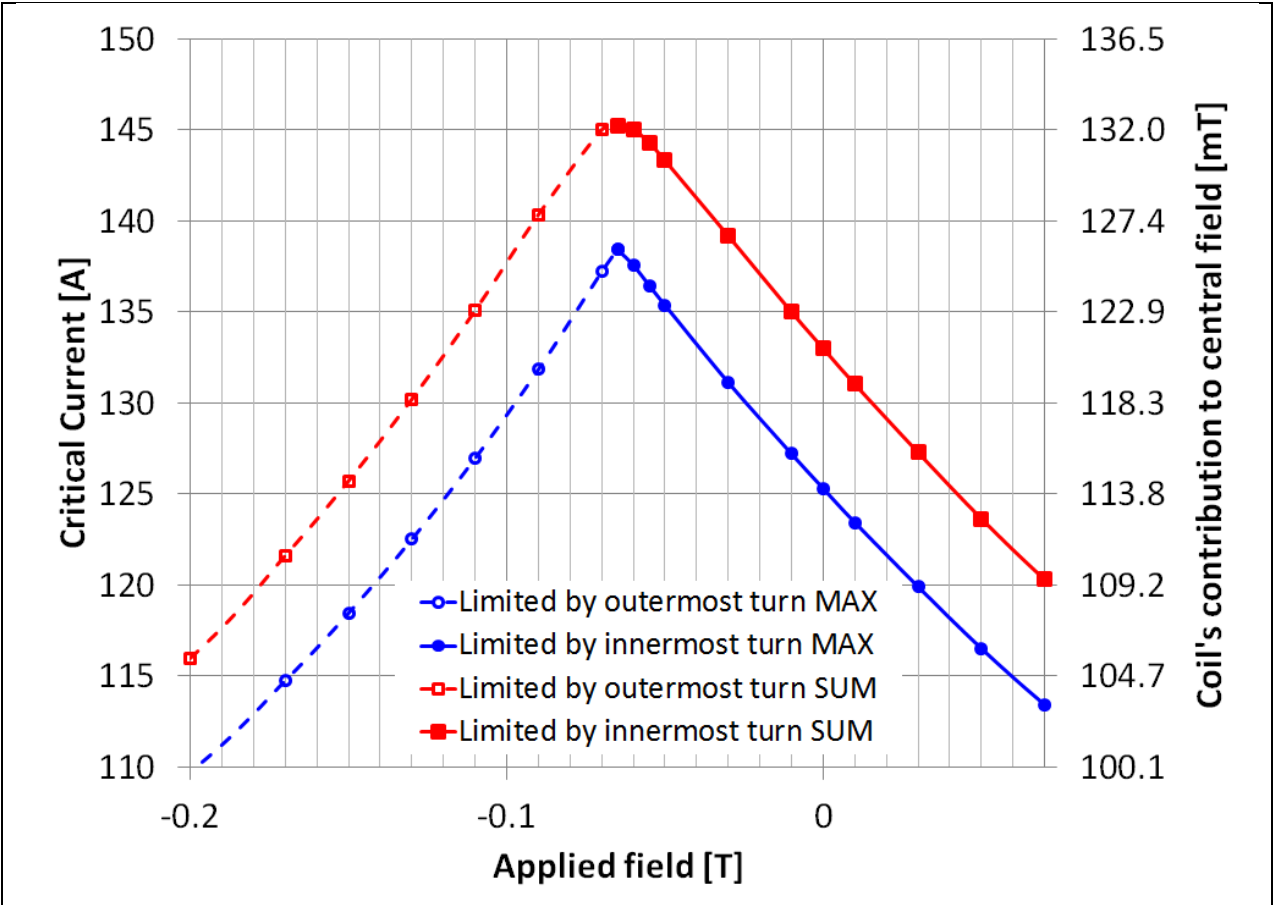

Figure 5 Critical current (left axis) and coils' contribution to central field (right axis) of a 50-turn pancake coil under externally applied field as estimated by both the MAX and SUM criteria.

%% This line is just a spacer %%

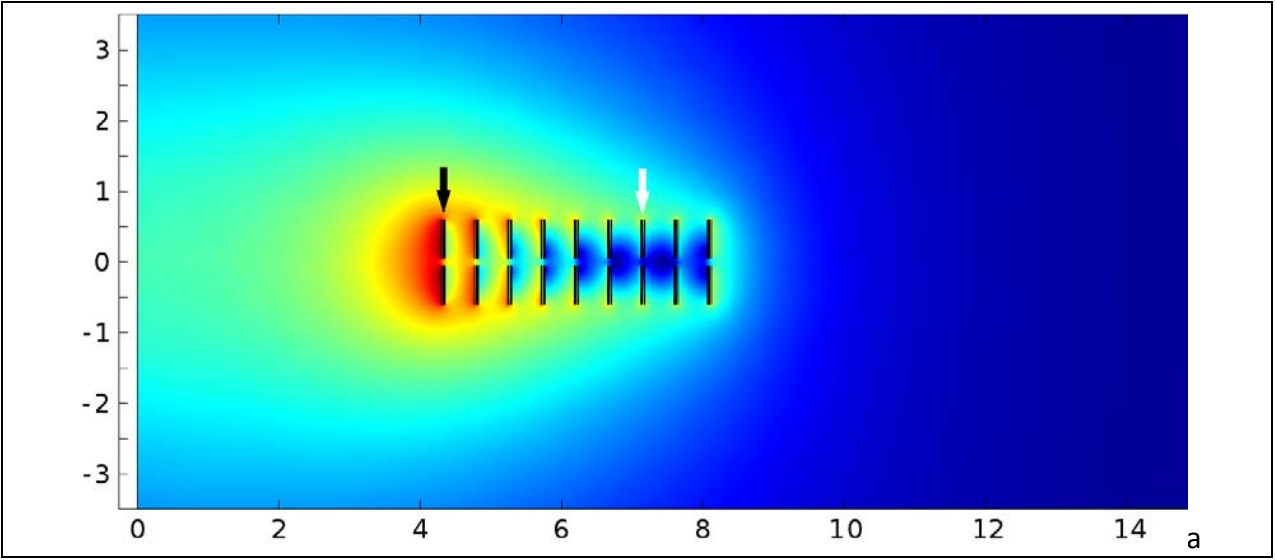

a

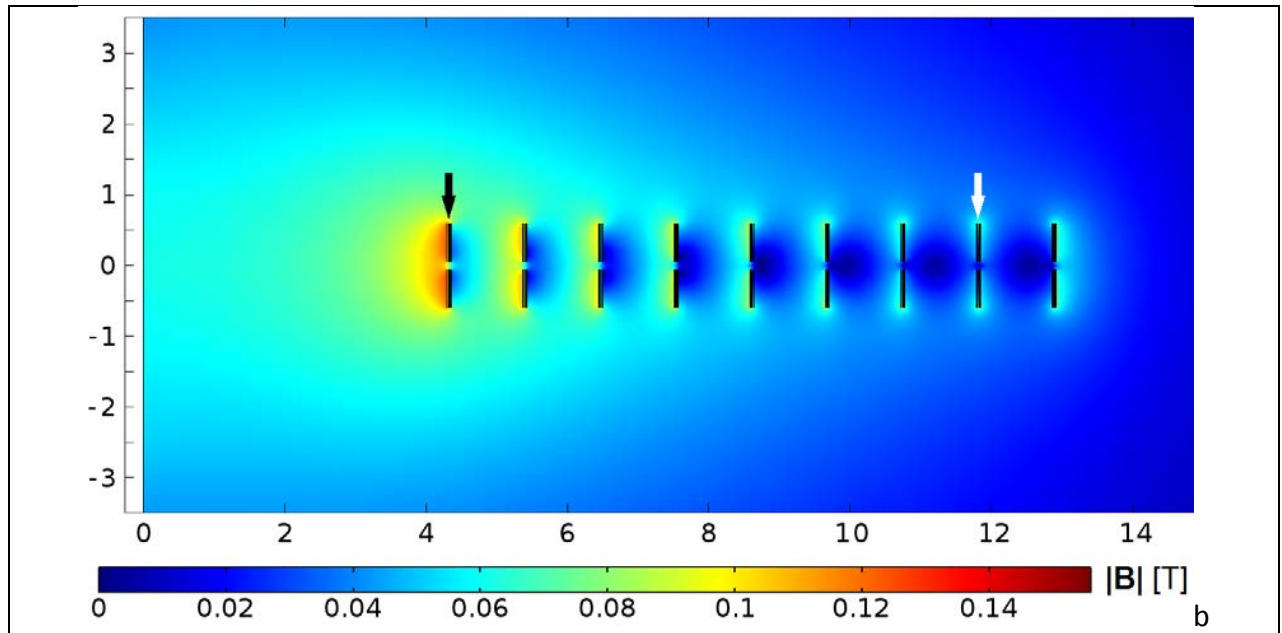

Figure 6. Magnetic flux density in half a cross section of Roebel coils with space between windings of 4mm (a) and 10mm (b). Both coils are at their critical current value given by the SUM-AVG criterion. Tick lines mark the geometric dimensions in cm. Considering the normalized current $\widehat{I}_\tau$ as defined in section 5.2, black and white arrows point respectively to the turn with the highest and lowest $\widehat{I}_\tau$ in each coil.

%% This line is just a spacer %%

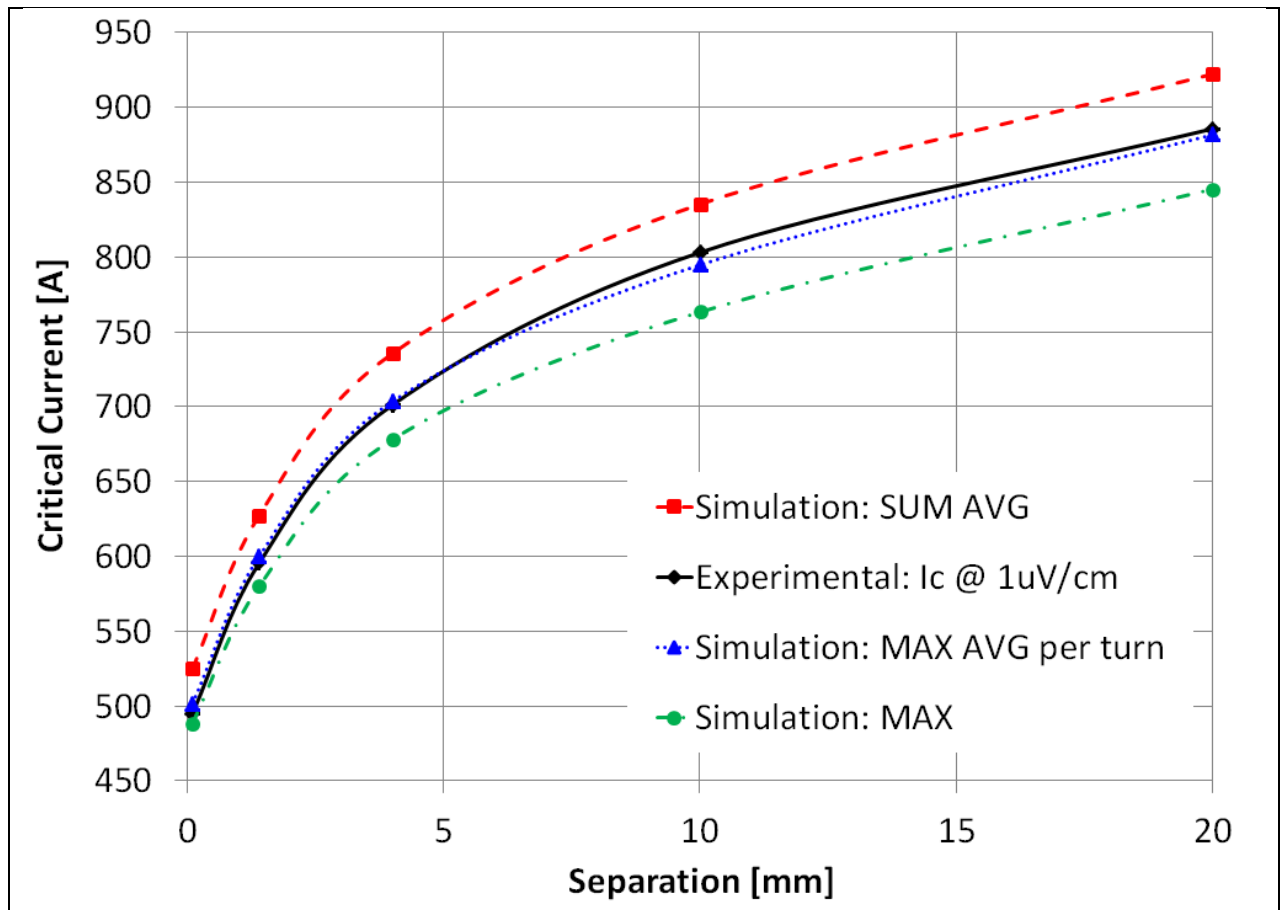
Figure 7 Critical current of Roebel coils with different inter-turn separation as estimated by the SUM-AVG, MAX-AVG and MAX criteria and as experimentally measured. The experimental setup corresponds to the SUM-AVG criterion.

%% This line is just a spacer %%

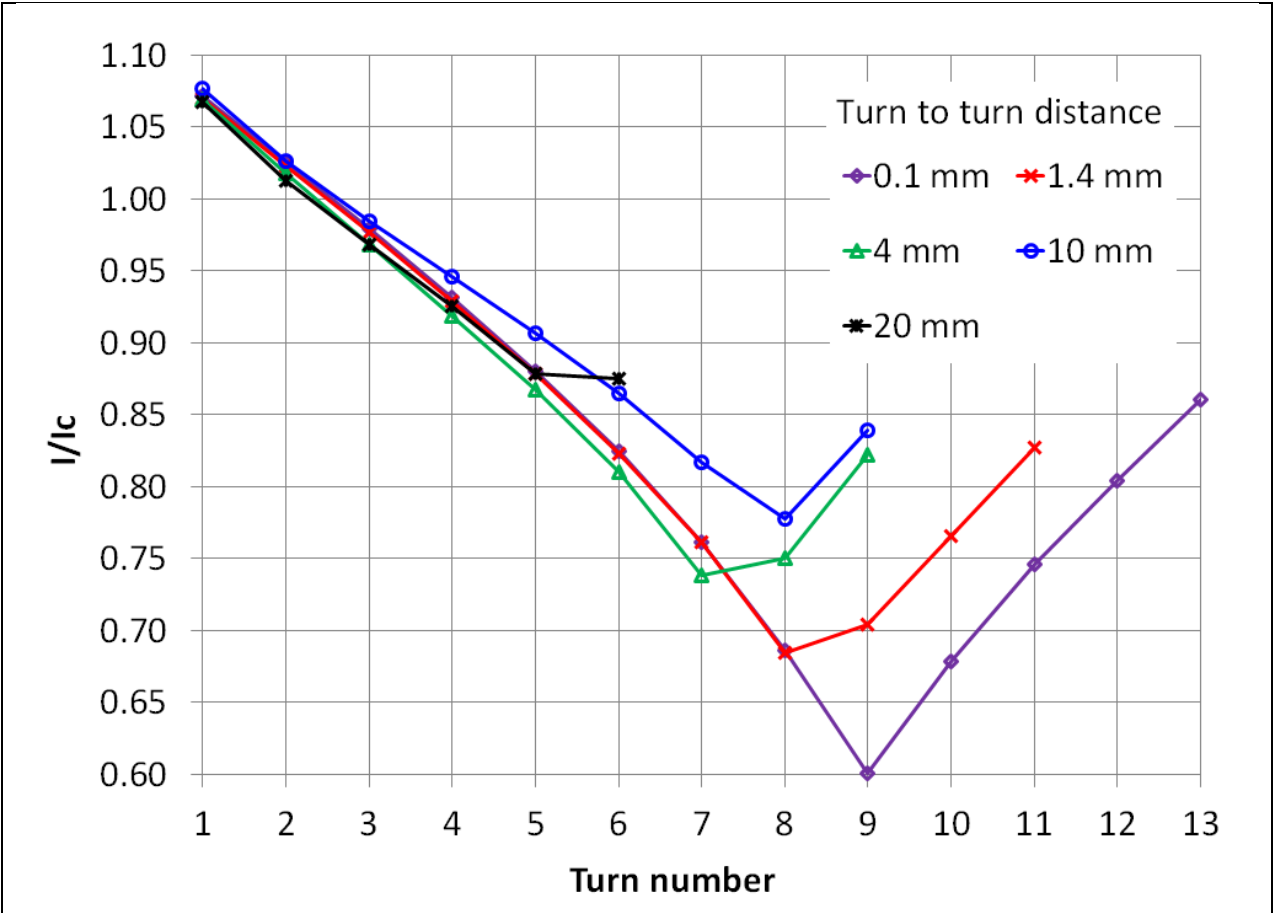

Figure 8 Normalized current $\widehat{I}_\tau$ in the turns of Roebel assembled coils as defined in section 5.2.

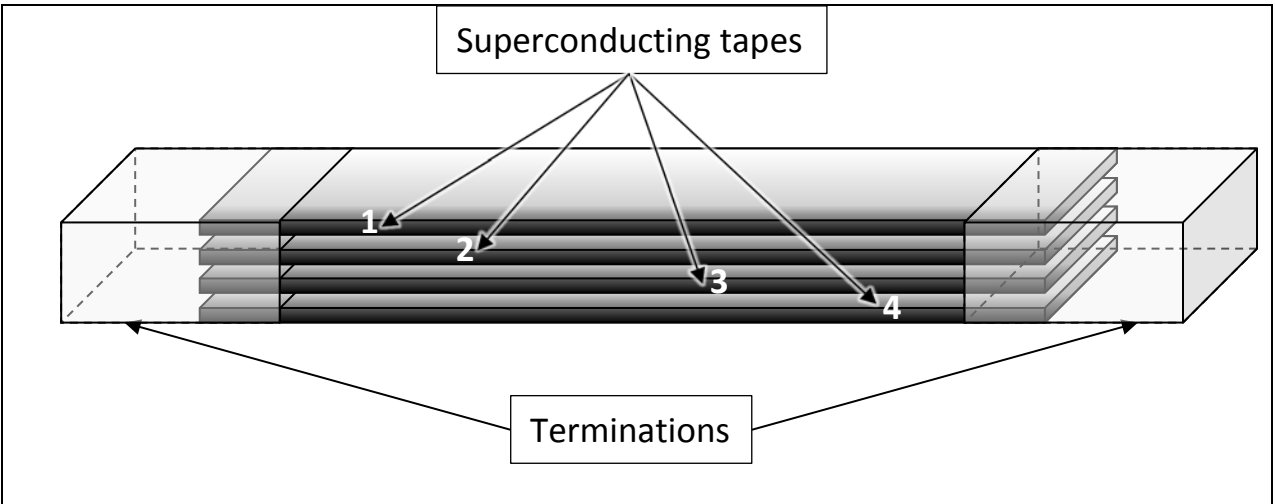

Figure 9. Layout of a stacked-tape cable (not to scale) composed of 4 HTS tapes. Ohmic terminations connect the tapes at both. Numbers 1 through 4 are used to denote the individual tapes in the cable.

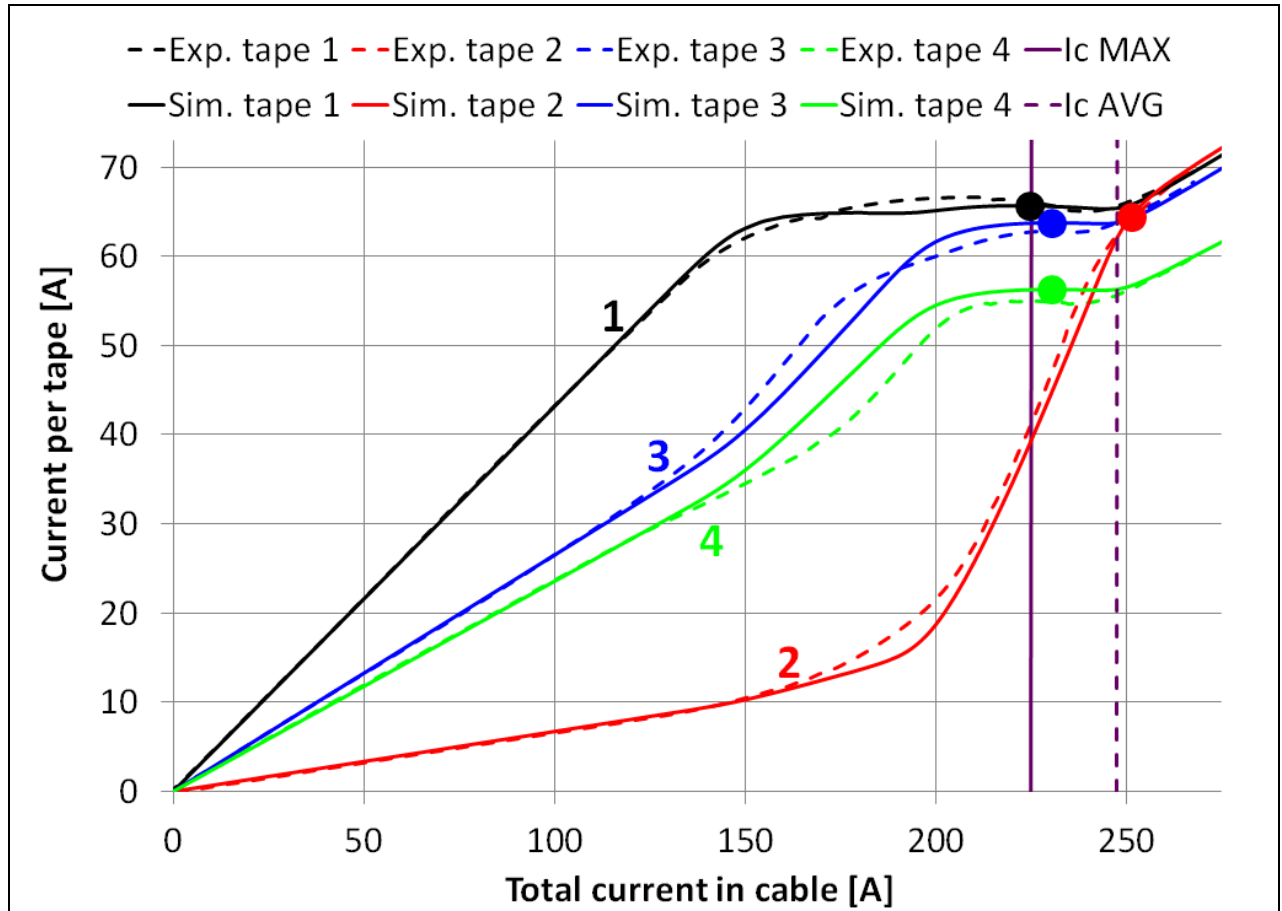

Figure 10. Current in individual tapes of a stacked tape cable as a function of the total current in the cable. For reference the critical current of the cable as estimated by the MAX and AVG criteria are marked respectively with solid and dashed purple vertical lines. The large dots indicate the total current in cable at which the individual critical current values of each tape while *in cable* configuration is reached.

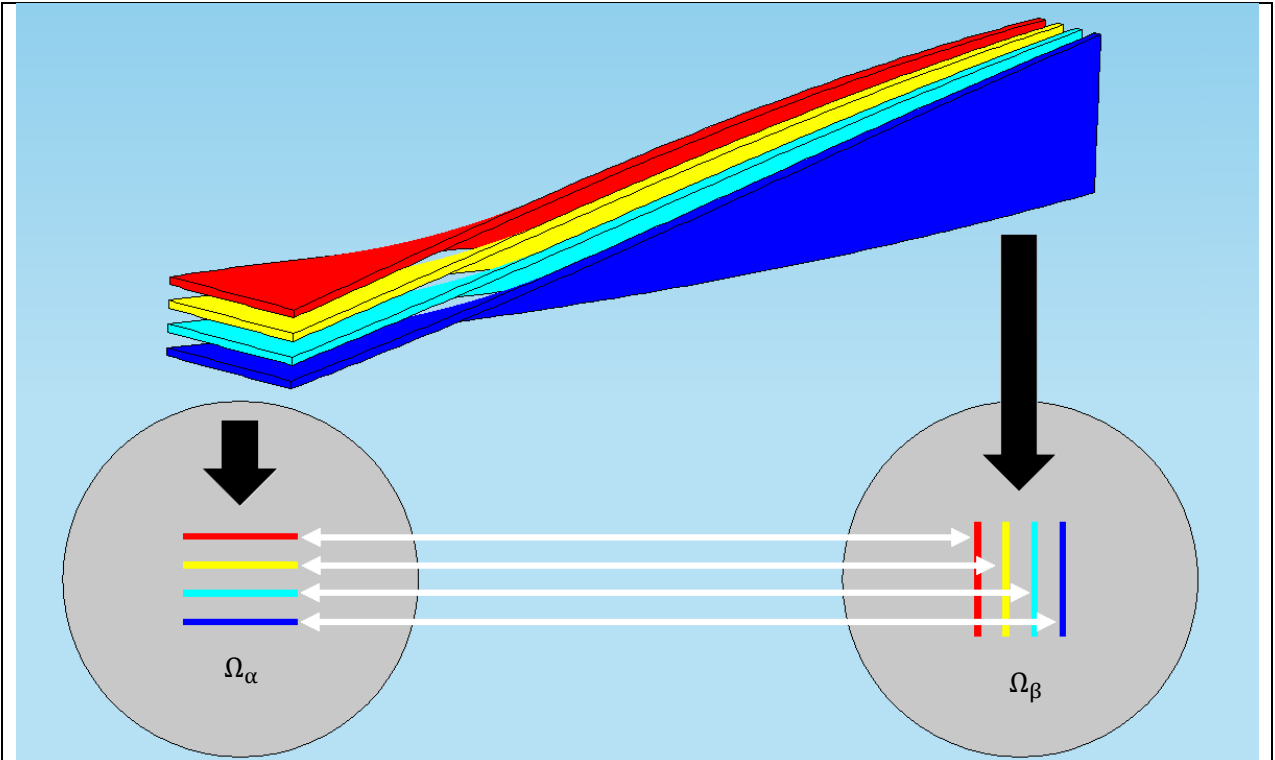

Figure 11. Schematic layout of the TSTC (top) and 2 slices used as computational domains (bottom).

%% This line is just a spacer %%

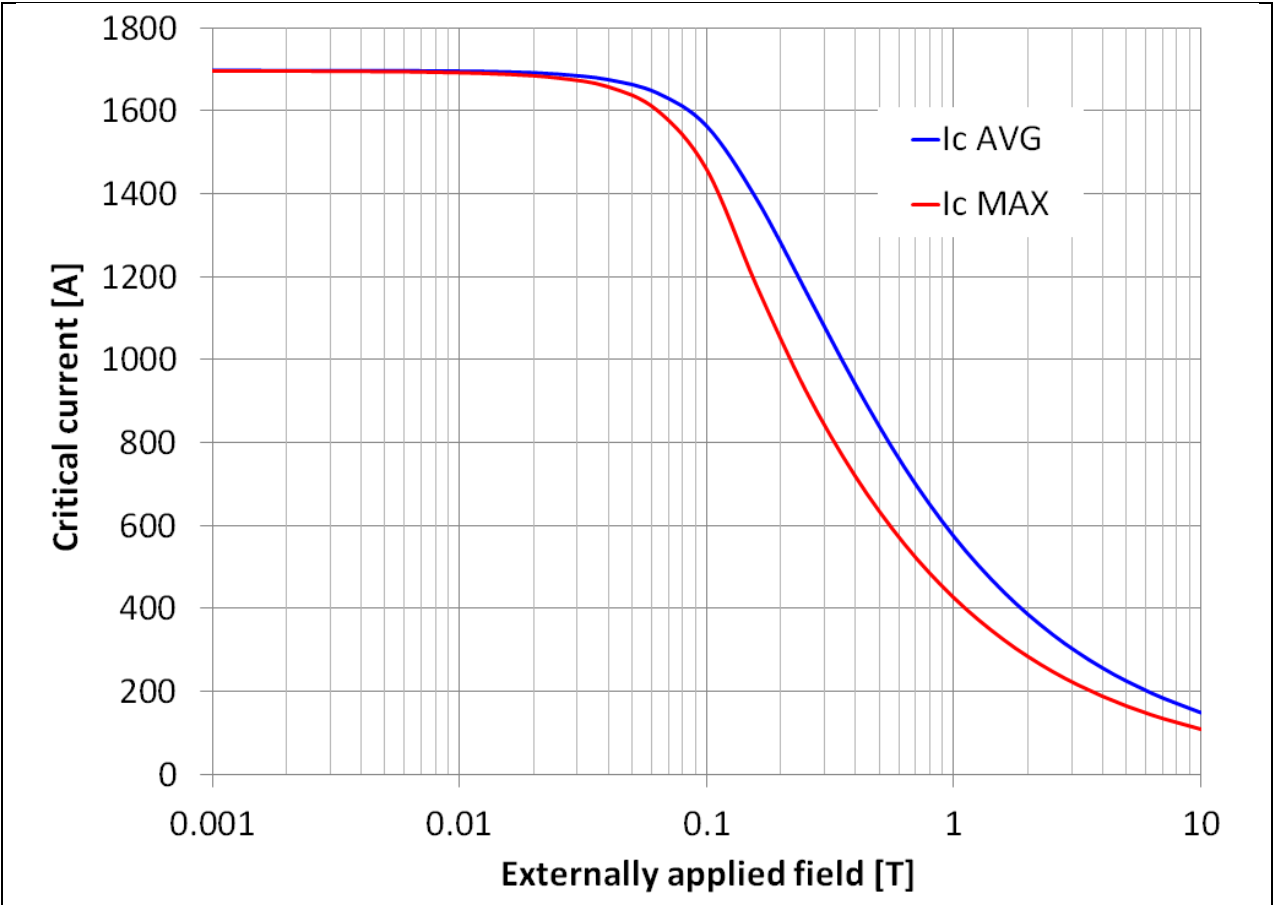

Figure 12 Critical current in a 40 tape TSTC under externally applied uniform magnetic field as estimated by both the AVG and MAX criteria.